\documentclass[prb,twocolumn,showpacs,preprintnumbers]{revtex4}
\usepackage{amsmath}
\usepackage{graphicx, epsfig}
\usepackage{dcolumn}

\def\be{\begin{equation}}
\def\ee{\end{equation}}
\def\bea{\begin{eqnarray}}
\def\eea{\end{eqnarray}}
\begin{document}

\title{ Thermally stimulated H emission and diffusion in hydrogenated amorphous silicon}
\author{T. A. Abtew}
\email{abtew@phy.ohiou.edu}
\affiliation{Department of Physics and Astronomy, Ohio University, Athens, OH 45701}
\author{F. Inam}
\email{Inam@phy.ohiou.edu}
\affiliation{Department of Physics and Astronomy, Ohio University, Athens, OH 45701}
\author{D. A. Drabold}
\email{drabold@ohio.edu}
\affiliation{Department of Physics and Astronomy, Ohio University, Athens, OH 45701}

\pacs{66.30.Jt,66.30.-h,61.43.Dq}

\begin{abstract}
We report first principles {\it ab initio} density functional calculations of hydrogen dynamics in hydrogenated amorphous silicon. Thermal motion of the host Si atoms drives H diffusion, as we demonstrate by direct simulation and explain with simple models. Si-Si bond centers and Si ring centers are local energy minima as expected. We also describe a new mechanism for breaking Si-H bonds to release free atomic H into the network: a fluctuation bond center detachment (FBCD) assisted diffusion. H dynamics in {\it a-Si:H} is dominated by structural fluctuations intrinsic to the amorphous phase not present in the crystal.
\end{abstract}
\maketitle

\section{Introduction} 

Hydrogenated amorphous silicon is one of the most important electronic materials, and is used in applications ranging from TFTs in laptop displays to solar photovoltaics and IR imaging/detection. Hydrogen dynamics is key to creation and annihilation of defects, and is also linked to light induced degradation \cite{fedders92,jackson88,branz99,abtew06}. Although there have been several studies, a complete picture of H dynamics is not yet available.

A widely held picture posits that unbonded H hops among various attractive sites before capture at a dangling bond \cite
{street91,santos92,vandewalle95,colombo96,biswas98,franz98}.
An intermediate low-energy pathway involving a metastable dihydride structure has also been reported \cite{tuttle99}. However, by calculating hopping rates for different trap sites, Fedders argued that thermal motion of hydrogen does not proceed from dangling bond to dangling bond via bond center (BC) sites and showed diffusion through intermediate levels to be insignificant \cite{fedders00}.   In addition, Su {\it et al} \cite{su02} proposed that Si-H bonds do not spontaneously release H, but rather require the mediation of an external agency: in their case a five-fold or ``floating" bond.
In crystalline Si, the importance of lattice dynamic activated diffusion has been reported \cite{panzarini,bedard}. Buda {\it et al.} \cite{buda} has shown diffusion of H in the form of jumps from BC site to another BC via intermediate hexagonal or tetrahedral sites.

Previous work on a-Si and a-Si:H showed that the network dynamics is in some ways quite different from the crystal. In particular, the disorder of the network allows fluctuations in the positions of atoms leading to the interesting observation of ``coordination fluctuation". It has been observed in an early first principles simulation that even at T=300K, the number of floating (fivefold) bonds fluctuated between zero and ten in a 216 atom cell, in a 1.8 ps simulation\cite{dad91}. This work has been updated, and similar effects have been observed in networks including H. It was also found that most of the atoms in the lattive eventually participated in these fluctuations\cite{springer}. 

In this Letter, we report an {\it ab initio} simulation which reveals the key role of thermal motion of Si atoms in driving H diffusion. We have undertaken accurate simulations including static lattice simulation (in which Si atoms were frozen) and extended thermal simulation. The static Si lattice simulations shows no H diffusion as compared with the dynamic lattice case, suggesting that the motion of the ``Si-sublattice" is important to the H dynamics. A key feature of our work is that we determine diffusion mechanisms directly from thermal MD simulation, {\it not} by imposing a conventional hopping picture among wells (traps) with varying depths. The principal conclusion of the paper is that the dynamic lattice (particularly the motion of pairs or triplets of Si atoms with a BC H present) is a primary means for ejecting atomic H into the network. This mechanism could not be easily inferred from phenomenological kinetic equation models of H transport \cite{fedders00}, though it should readily emerge using a method devised to discover rare (long time scale) events like the Activation-Relaxation Technique implemented with {\it ab initio} interactions \cite{wales}.

We determine an essential mechanism for H diffusion in the dynamic lattice, namely, ``Fluctuating Bond Center Detachment" (FBCD) assisted diffusion: if the H is initially covalently bonded to a Si atom, it stays bonded with it until another Si comes in the vicinity and makes an  instantaneous or fluctuating BC configuration. This event is followed by a switching of H from the covalent bond to the new Si to either form another Si-H bond or hop, depending upon the local environment. This process is important both as a means for the network to generate free H and to create dangling bonds. The mechanism we report here is undoubtedly not the sole means of obtaining H diffusion, but is predominant in accurate and relatively extended MD simulations. 

\section{Methods}

\begin{figure}
\begin{center}
\includegraphics[width=0.45\textwidth]{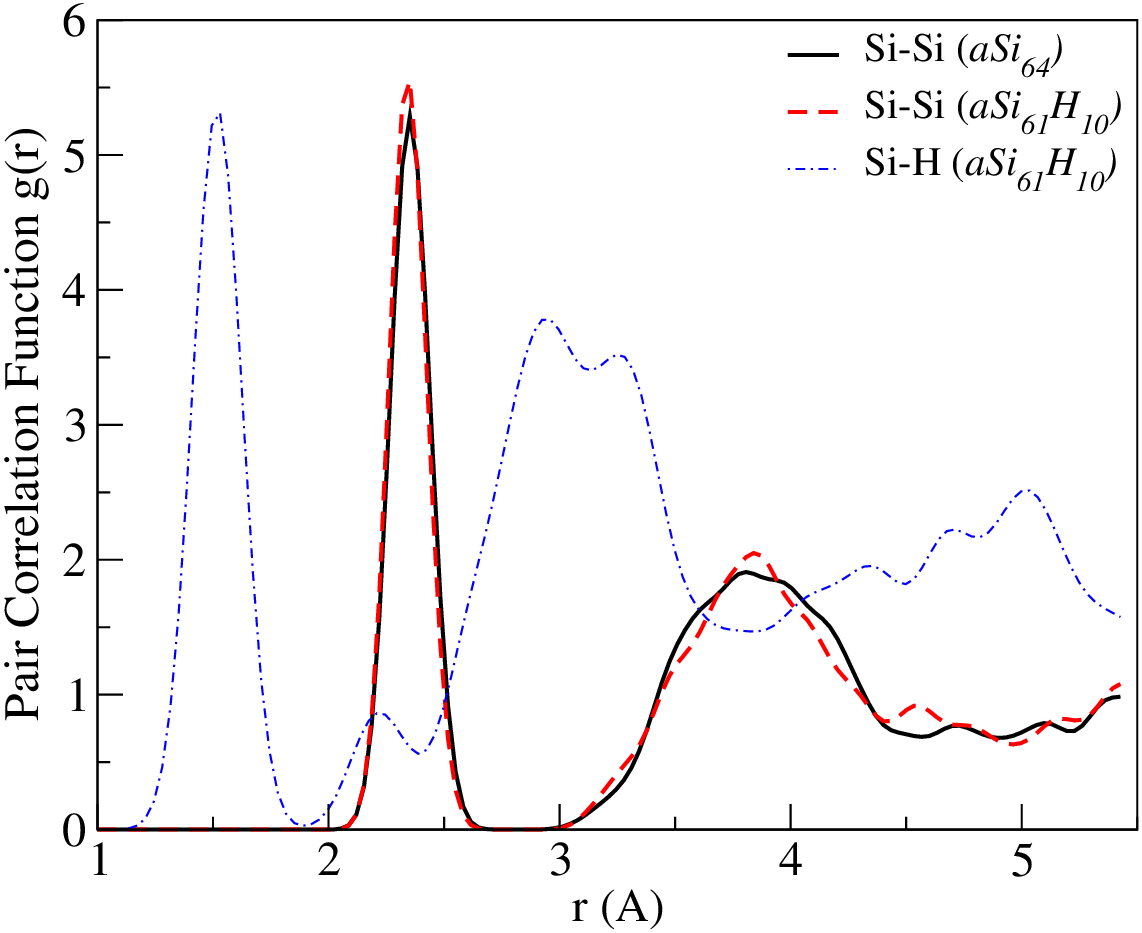}
\end{center}
\caption{\label{gor}Si-Si and Si-H partial pair correlation for {\it aSi$_{61}$H$_{10}$}} and Si-Si partial pair correlation for {\it aSi$_{64}$} models.
\end{figure}

To understand the diffusion of H at different temperatures we have used a 71 atom model {\it aSi$_{61}$H$_{10}$}. We generated this model by removing three Si atoms from a well relaxed defect free {\it aSi$_{64}$} model~\cite{mousseau}. We then terminated all the dangling bonds except two with hydrogen. The newly formed supercell is then relaxed using conjugate gradient to form a {\it aSi$_{61}$H$_{10}$} model (61 Si and 10 H atoms). The computed Si-H partial correlation has the expected peak at 1.5~\AA~ and the Si-Si partial pair correlation show a first peak at 2.35~\AA~with a subsequent minimum which remains unchanged with hydrogenation as shown in Fig.~\ref{gor}. The newly generated model retains a tetrahedral structure and fourfold coordination (for Si) as in the original.
We have also considered additional models, a 138 atom model {\it aSi$_{120}$H$_{18}$}~\cite{fedders} and a 223 atom model {\it aSi$_{214}$H$_{9}$}~\cite{www,223model}, and studied H dynamics there to assure ourselves that finite-size effects were unimportant.

The simulations were performed using SIESTA~\cite{soler} in the generalized gradient approximation for the exchange (GGA) using a parametrization of Perdew, Berke, and Ernzerhof~\cite{pbe}. Norm conserving Troullier-Martins pseudopotentials~\cite{tm} factorized in the Kleinman-Bylander form~\cite{kb} were used. We used a single $\zeta$ polarized basis set for Si valence electrons and double $\zeta$ polarized basis for H allowing sufficient flexibility in the basis set. We solved the self-consistent Kohn-Sham equations by direct diagonalization of the Hamiltonian and a conventional mixing scheme. Every structure in this report was relaxed using conjugate gradient (CG) coordinate optimization until the forces on each atom were less than 0.02 eV/\AA. The $\Gamma$ point was used to sample the Brillouin zone in all calculations.

\section{Simulations}

Our strategy has been to study the dynamics of H atoms in small cells which appear to represent the local topology of the amorphous network, using highly accurate methods. The use of such methods precludes simulations much exceeding several picoseconds, particularly because of the small mass of H and the commensurately small time step needed to integrate the equations of motion.

\begin{figure}[htbp!]
\begin{center}
\includegraphics[width=0.45\textwidth]{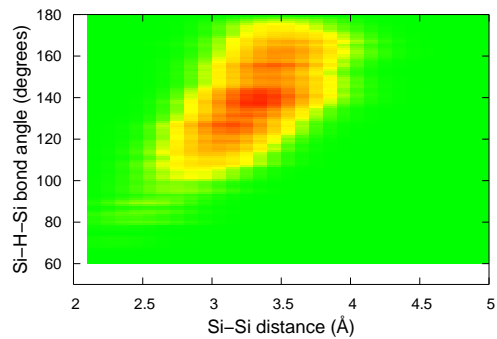}
\caption{\label{avertheta} Dynamics of H near BC conformations. The contours give normalized temporal distribution of R$_{\textrm {Si-Si}}$ as a function of Si-H-Si bond angle from aSi$_{61}$H$_{10}$ for a total of 6.25 ps at T=1000K: red, most frequent visitation and green least frequent. The plot is averaged over all H atoms in the cell when in a BC conformation.}
\end{center}
\end{figure}

In our simulations, H in a-Si:H is either passivating Si dangling bonds, or occupying other sites, such as BC sites (the most common) or certain other locations that provide energy minima for H. To provide some information about the dynamics of our cells, we have considered dynamics near BC conformations in aSi$_{61}$H$_{10}$.  In Figure~\ref{avertheta}, we have plotted,
$ \rho(R,\theta) \propto \sum_i\sum_j\delta(R-R_i(t)) \delta(\theta-\theta_j(t))$,
 a distribution function indicating time spent in different parts of the $R-\theta$ configuration space, where $R_i(t)$ is the distance between two Si atoms at a time, and $\theta_j(t)$ is the (Si-H-Si) angle formed by two silicons and a central H at a given time.  This result shows preferred values of $R$, ranging from 3.0 - 4.0 \AA~and $\theta$ (Si-H-Si bond angle) in the range of 110$^\circ$ - 180$^\circ$ where the H atoms spends most time. The hydrogen trapping time is highest on the red and lowest on the green regions. Evidently, the energy surface is rather weakly dependent upon $\theta$ over a wide range of angles.

To analyze the role of thermally induced Si motion in driving H diffusion, we considered 6.25 ps simulations (25 000 MD steps each with time step $\tau$=0.25 fs) for {\it aSi$_{61}$H$_{10}$} at T=300K and T=1000K. The difference in the H diffusion for these two cases is easily extracted from the time-averaged mean squared displacement of H, which is defined as
$\langle \sigma^2 \rangle_{\textrm {time}}=\frac{1}{{N_{\tiny MD}N_ {\tiny H}}} \sum_{t=1}^{N_{\small MD}} \sum_{i=1}^{N_H}  |\vec{r}_i(t)-\vec{r}_i(0)|^2$, where $N_{\small MD}$ is the total number of MD step, $N_H$ and $\vec{r}_i(t)$ are total number and coordinates of H at time $t$ respectively. This yielded an average mean square displacement of 0.14 \AA$^2$ for T=300K and 1.06 \AA$^2$ for T=1000K. Similar analysis on Si gives an average mean square displacement of 0.06 \AA$^2$ and 0.57 \AA$^2$ for T=300K and T=1000K respectively. 



\section{Discussion}

In the case of  {\it aSi$_{61}$H$_{10}$} at T=1000K, we have
observed 9 bond breaking events which are accompanied by rapid bond
switching in the 6.25 ps simulation time. For {\it aSi$_{214}$H$_9$}
at 300K we have observed 5 bond breaking events in the 10 ps
simulation time; all are FBCD assisted.  Only one bond breaking
event we observed for {\it aSi$_{61}$H$_{10}$} at 300K for the total
simulation time of 6.25 ps. We observed no bond breaking event for
{\it aSi$_{120}$H$_{18}$} model at 300K, 500K and 700K temperatures.
After the bond breaking, all of the events lead to the hydrogen
passivating a dangling bond. To discuss the FBCD mechanism in
detail,  we have selected two H from {\it aSi$_{61}$H$_{10}$} model
namely H$_{67}$ and H$_{68}$ at T=1000K, which diffuse through via
FBCD. To analyze the role of the thermal motion of the neighboring
Si atoms we tracked all nearby Si pairs correlated with the motion
of H in both cases.

\begin{figure}[htbp!]
\begin{center}
\includegraphics[ width=0.45\textwidth]{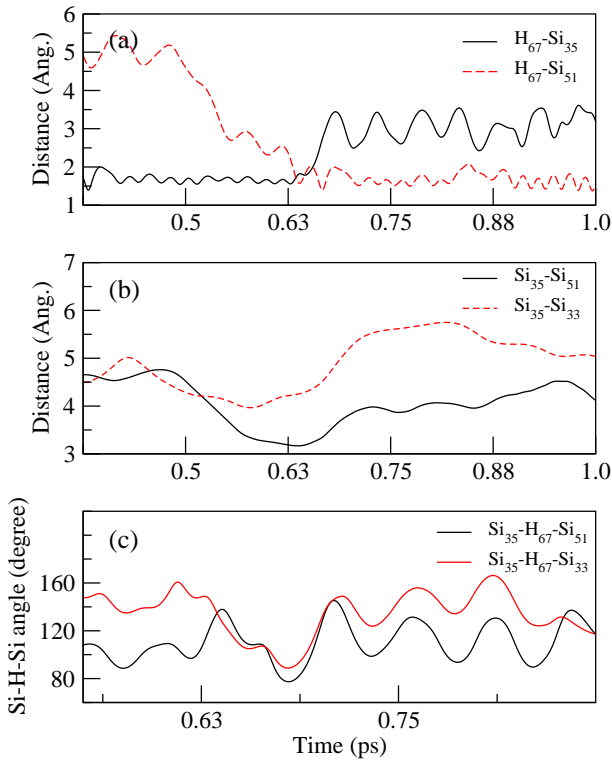}
\end{center}
\caption{\label{H-H67} H$_{67}$ switching the bond from Si$_{35}$ to
Si$_{51}$ assisted by the close approach of Si$_{33}$. a) shows the
swhiching of H bond between the two Si. b) and c) shows the
fluctuation of distances of Si$_{33}$ and Si$_{51}$ from Si$_{35}$
and variation in the corresponding angles centered at H$_{67}$
respectively.}
\end{figure}

In Figure~\ref{H-H67}, we show a situation in which H$_{67}$ initially bonded to Si$_{35}$ switched to Si$_{51}$. This event follows the close approach of Si$_{51}$ and Si$_{33}$ to the Si$_{35}$-H$_{67}$ bond and forms a fluctuating bond center conformation. As the distance Si$_{35}$-Si$_{51}$ and Si$_{35}$-Si$_{33}$ change from 4.8~\AA~to 3.1~\AA~and from 5.0~\AA~to 4.0~\AA~respectively as shown in Fig.~\ref{H-H67}(b), the bond angle Si$_{35}$-H$_{67}$-Si$_{51}$ changes from 140$^\circ$ to 80$^\circ$. As the same time, the bond angle Si$_{35}$-H$_{67}$-Si$_{33}$ also changes from 160$^\circ$ to 95$^\circ$.  These introduce change from a BC configuration, which is a local energy minima, to energetically unstable part of the $R - \theta$ configuration space, compelling the H to diffuse and form a bond with another Si.

\begin{figure}[htbp!]
\begin{center}
\includegraphics[ width=0.45\textwidth]{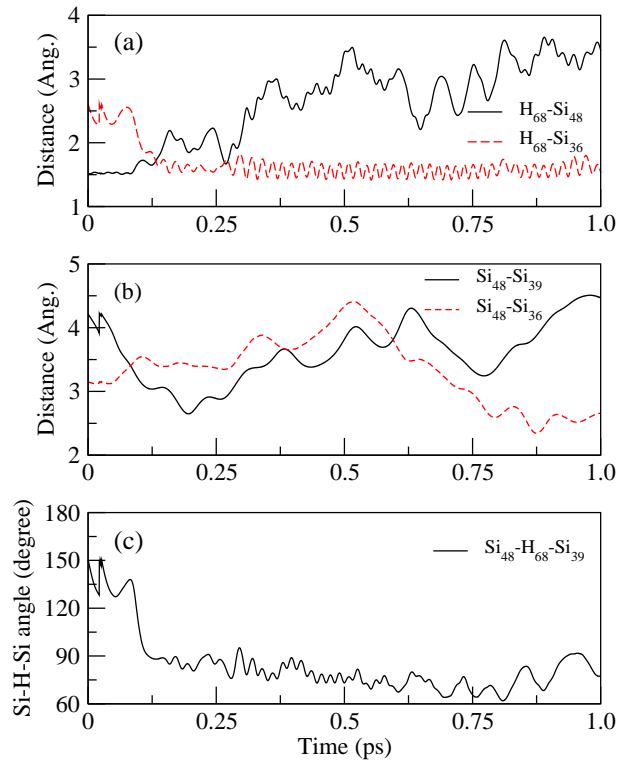}
\end{center}
\caption{\label{H-H68} Another H Bond switching event. a) shows
H$_{68}$ switching bond from Si$_{48}$ to Si$_{36}$. b) and c) shows
the fluctuation in the distances of Si$_{36}$ and Si$_{39}$ from
Si$_{48}$ and variation in the corresponding Si-H-Si angles
respectively.}
\end{figure}

A second case for the FBCD assisted diffusion involves H$_{68}$ similar to the previous example. As shown in Fig.~\ref{H-H68}(a), we observed a bond breaking process in H$_{68}$, a situation where H$_{68}$ which was initially bonded to Si$_{48}$ and later switched to Si$_{36}$. The switching occurs due to the close approach of Si$_{39}$ to the Si$_{48}$-H$_{68}$ bond and forms a fluctuating bond center conformation. As the distance Si$_{48}$-Si$_{39}$ changes from 4.2~\AA~to 3.0~\AA~as shown in Fig.~\ref{H-H68}(b), the bond angle Si$_{48}$-H$_{68}$-Si$_{39}$ changes from a bond center configuration 150$^\circ$ to 89$^\circ$, compelling the H to diffuse and form a bond with another Si.

\begin{figure}[htbp!]
\begin{center}
\includegraphics[ width=0.45\textwidth]{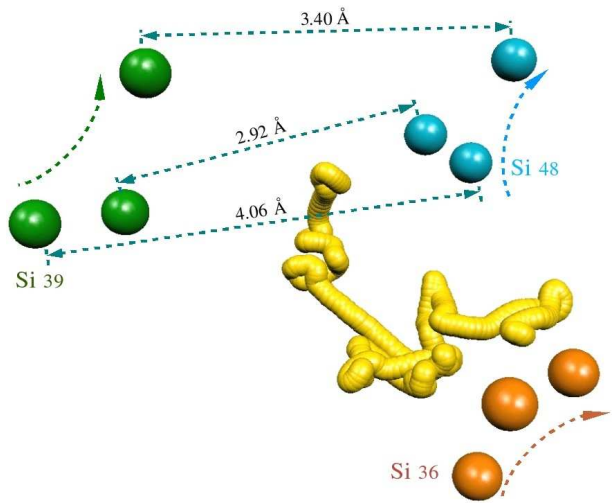}
\end{center}
\caption{\label{H68-new2} Fluctuating Bond Center Detachment and diffusion. Trajectory (averaged over 400 steps) of H$_{68}$ are shown in yellow, while Si$_{48}$, Si$_{36}$, and Si$_{39}$ are shown with blue, orange, and green colors respectively for three different time steps of the simulation. }
\end{figure}

In Fig. \ref{H68-new2}, we show the trajectory of  H$_{68}$ in FBCD assisted diffusion. The average positions of H$_{68}$ are shown in yellow, while Si$_{48}$, Si$_{36}$, and Si$_{39}$ are shown with blue, orange, and green colors respectively at three different time steps of the simulation. H$_{68}$ is initially bonded to Si$_{48}$, becomes a fluctuating bond center H with approach of Si$_{39}$, is ejected and eventually bonds to Si$_{36}$.

The FBCD mechanism is reminiscent of that of Su {\it et al.} \cite{su02}, also depending upon the intercession of a Si not part of the initial conformation. However, the FBCD is a more general process that may or may not increase the Si coordination. We found many cases of H detachment in which no Si was overcoordinated. Ejection of H is a more subtle process than just changing coordination, and depends on the the local geometry (R, $\theta$). Finally, all such FBCD conformations arise from fluctuations, and are thus short lived \cite{dad91}. The mechanism of Su {\it et al} is a special case of FBCD.

In conclusion, we have demonstrated the nature of H diffusion in a-Si:H by direct simulation and with the aid of a model to develop a fairly simple picture of H motion. H emission is stimulated by Si motion, and the FBCD mechanism is shown to be important both for stripping off H chemically bonded to Si (thus creating ``free" atomic H), and of course for creating Si dangling bonds. Our work is consistent with analogous studies in c-Si \cite{buda}, and is a generalization of the work of Su and Pantiledes \cite{su02}.

\section{Acknowledgments}
We thank the National Science Foundation for support under grant No. DMR 0600073 and the Army Research Office under MURI W91NF-06-2-0026. We acknowledge helpful conversations with P. C. Taylor, E. A. Schiff and P. A. Fedders.  Some of this work was carried out when DAD visited the Institut de Ciencia de Materials de Barcelona with support from the Programa de Movilidad de Investigadores of Minesterio de Educacion y Cultura of Spain.

\end{document}